\documentstyle[12pt,prepriop]{article}
\begin{document}
\begin{center}
\bf   THE DYNAMICS OF VORTEX STRUCTURES AND STATES OF CURRENT IN     
PLASMA-LIKE FLUIDS AND THE ELECTRICAL EXPLOSION OF CONDUCTORS:\\
2. Computer experiment.
\end{center}
\begin{center}
N B Volkov and A M Iskoldsky\\
Russian Academy of Science, Ural Division\\
Institute of Electrophysics\\
34 Komsomolskaya St.\\
Ekaterinburg 620219,\\ 
Russia
\end{center}
\vspace{3.5cm}
\begin{center}
\bf The Dynamics of Vortex Structures and States of Current: 2
\end{center}
\newpage
\begin{center}
{\bf THE DYNAMICS OF VORTEX STRUCTURES AND STATES OF CURRENT IN 
    PLASMA-LIKE FLUIDS AND THE ELECTRICAL EXPLOSION OF CONDUCTORS:\\    
2. Computer experiment.}\\ 
                     N B Volkov and A M Iskoldsky
\end{center}
\begin{center}
\bf Abstract
\end{center}
\par
In the present paper which is a sequel to 
[N~B~Volkov and A~M~Iskoldsky The dynamics of vortex structures 
and states of  current:~1;[1]], 
the dynamics of non-equilibrium phase transitions and states  of
current in electrophysical systems providing an external circuit and a
nonlinear element the model of which has been developed 
in [1] (in the
development of the model local  kinetic  transport  coefficients  were
assumed to be constant), has been analyzed and simulated. A  non-equilibrium
phase transition has been shown to be induced by large-scale 
hydrodynamic fluctuations (vortex  structures).  
Critical  exponents  of  the
amplitude singular behavior (the order parameters) for three types  of
circuits have been determined. Non-equilibrium phase 
transitions induced by external harmonic noise 
with a random or a determined  phase
have been studied. The detuning between an  external  noise  frequency
and  natural  frequencies  of  a  nonlinear  element   determined   by
large-scale hydrodynamic fluctuations have been shown to give rise  to
random oscillations, which are typical of a  strange  attractor.  When
the frequencies coincide, a limit cycle appears in the  system,  which
is characterized by the fact that the phase trajectory does  not  fill
the phase space completely.
\newpage
\section{Introduction}
\par
In the first paper of this series [1], 
a dynamic  model 
of a non-equilibrium phase  transition  (NPT)  has  been  offered  and
studied. This model, in our opinion, describes the  initial  stage  of
the turbulence nucleation in a conducting fluid  (see  our  
paper [2] 
which indicates an analogy between the initial stage of the electrical
explosion of a conductor (EEC)  and  the  turbulence  nucleation  in  an
incompressible liquid. It was also noted in [1] 
that a further evolution of  vortex  structures, 
corresponding  to  a  transition  to  an 
isotropic spectrum which is characteristic of a developed 
turbulence, 
would be a consequence of the splitting of the  space scale. 
Assuming, 
that the splitting is absent or hindered (as follows from 
[1], for the 
splitting to take place, the existence of a laminar flow of  fluid  is
essential), then already  in  the  framework  of  model 
[1],  general 
problems of the dynamics of states of current in a  conducting  liquid
can be studied, particularly, the dynamics of a transition to a steady
state of the turbulence resistance. The model of 
[1] is a model of a 
nonlinear element (NE) inserted into the electric circuit that 
is a thermostat with respect to the NE. 
\par
The purpose of the present computer experiment was to study 
proper motions that appear in the dynamic system, 
consisting  of  the NE
model and the electric circuit; the dynamics of a spontaneous breaking
symmetry in the initially laminar electron fluid,  and  also  the  NPT
dynamics, induced by either a determinate  or  a  stochastic  external
noise.
\section{Equations. Procedure.}
\par
We discuss three types of an external electric circuit: 
S1 (Fig. 1a), S2 (Fig. 1b), S3 (Fig. 1c), where 
$L_0$ is the inductance, $C_0$  is  the 
capacitance, $R_1$ is the ballast resistance, $R_2$ and 
$L_{p0}$ are ,respectively, the initial resistance and 
the NE inductance, $R_3$  is  the  load 
impedance; $I_1$, $I_2$ and $I_3$, are the currents 
in the  corresponding branches of a network. Circuit S1 can be 
easily transformed to the DC current source, if we formally 
let $L_0\rightarrow\infty$. Similarly, 
circuits S2 and S3 can be easily transformed 
to  the DC voltage sources, if we formally let 
$C_0\rightarrow\infty$.
\par
Equations describing transient processes in the circuits are: 
\par
{\bf A. The model of a nonlinear element:}
\begin{eqnarray}
{\dot X} & = & s(-X+{I_2}Y),         \\
{\dot Y} & = & {a_1}X(-Z+{a_1}{r_1}{I_2})-Y,   \\
{\dot Z} & = & -({a_1}XY+bZ),
\end{eqnarray} 
where the point symbol $"\cdot"$ stands for the differentiation operator 
of a dimensionless time $\tau = t{t_0}^{-1}$, 
$t_0 = (4{g_1}^2{\nu_m})^{-1}b{r_0}^2$ is the basis time; 
$I_2 ={i_2}{I_0}^{-1}$ is the dimensionless current; $r_1 =
R{R_c}^{-1}$ is the control  parameter 
of the model; $a_1 = \pi{g_1}^{-1} = 0.819$; 
$g_1 = 3.83171$ corresponds to the first zero of the Bessel 
function ${J_1}(x)$; $b=8/3$; $R$ and $R_c$ are, 
respectively, the magnetic
Rayleigh number and its critical value. 
\par
As it was noted in [1], knowing 
$X$($\tau$), $Y$($\tau$), $Z$($\tau$), and $I$($\tau$)  enables
determination of the paths of particles transferring 
mass ("hydrodynamic particles") and current 
(conduction  electrons),  while  solving the motion equations:
\begin{eqnarray}
{{\dot R}_a} & = & -{C_r}X{\cos ({\pi}k{Z_a})}{{J_1}({g_1}{R_a})}, \\
{{\dot Z}_a} & = & {C_z}X{\sin ({\pi}k{Z_a})}{{J_0}({g_1}{R_a})}
\end{eqnarray}       
and 
\begin{equation}
{{\dot R}_e} = {\sqrt{2}}Bk{r_1}^{-1}Y
{\sin ({\pi}k{Z_e})}{J_1}({g_1}{R_e}),
\end{equation}
\begin{eqnarray}
{{\dot Z}_e} & = & B(2{I_2}+({a_1}{r_1})^{-1}{\biggl(}{\sqrt{2}}Y
{\cos ({\pi}k{Z_e})}{{J_1}({g_1}{R_e})}+ \nonumber\\
& & +2Z({{J_1}^2({g_1}{R_e})}-{{J_0}^2({g_1}{R_e})}){\biggr)},
\end{eqnarray} 
where $R_a$, $Z_a$ and $R_e$, $Z_e$ are the 
coordinates of the paths for atoms and 
conduction electrons respectively; 
$C_r = 2^{1/2}{\pi}{g_1}^{-2}$; $C_z = {g_1}k^{-1}C_r$; 
$B = bcH_0(16{\pi}e{n_e}{\nu_m}{g_1}^2)^{-1}$; 
$e$ and $n_e$ are the unit 
charge  and  the 
conduction electron density respectively. Since Eqs.(4) - (7) 
determine the particle paths corresponding to each moment 
of time  $\tau$,  then 
in their integration the $X$, $Y$, $Z$ and $I$ amplitudes 
should  be  regarded as 
constant, i.e. sets (4) - (5) and (6) - (7) are autonomous 
dynamic systems of  the 
second order on the plane $(r,z)$. 
\par
{\bf B. Circuit S1:}
\begin{equation}
{{\dot I}_2} = -[{{\Pi}_1}{I_2} + {\Pi}_2(I_2 - a{r_1}^{-1}Z)],
\end{equation}
where ${\Pi_1} = {R_1}c^2{t_0}{L_0}^{-1}$; 
${\Pi_2} = {R_{p0}}{c^2}{t_0}{L_0}^{-1}$; 
$a = {\pi}^{-1}{J_0}^2(g_1)$. 
\par
{\bf C. Circuit S2:}
\begin{eqnarray}
{\dot U} = - {\Pi_4}I_1, \\
{\dot I}_2 = {G_2}^{-1}[U + {G_1}{\Pi_2}a{r_1}^{-1}Z], \\
{\dot I}_1 = (1 +{\Pi_2}{\Pi_3}^{-1})I_2,
\end{eqnarray}
where $\Pi_2 = R_{p0}{R_1}^{-1}$, $\Pi_3 = {R_3}{R_1}^{-1}$, 
$\Pi_4 = {t_0}({R_0}C_0)^{-1}$, 
$G_1 = 1 + {\Pi_3}^{-1}$, $G_2 = 1+ {G_1}\Pi_2$. 
\par
{\bf D. Circuit S3:}
\begin{eqnarray}
{\dot I}_1 = {\Pi_5}U - [{\Pi_1}I_1 + {\Pi_6}{\dot I}_2 + U_R], \\
\dot U = - {\Pi_4}I_1, \\
{\dot I}_2 = {\Pi_6}^{-1}[{\Pi_3}(I_1 - I_2) - U_R],
\end{eqnarray}
where $U_R = {\Pi_2}(I_2 - a{r_1}^{-1}Z)$; 
$\Pi_1 = {R_1}{c^2}{t_0}{L_0}^{-1}$; 
$\Pi_2 = R_{p0}{c^2}{t_0}{L_0}^{-1}$; 
$\Pi_3 = R_3{c^2}{t_0}{L_0}^{-1}$; 
$\Pi_4 = I_0{t_0}(U_0{C_0})^{-1}$; 
$\Pi_5 = U_0{c^2}{t_0}(I_0{L_0})^{-1}$;  
$\Pi_6 = L_{p0}{L_0}^{-1}$; 
$U_0$, $I_0$ are the basis voltage and the current. 
If we choose $U_0$  and $I_0$ 
so that $U_0 = {R_1}I_0$, then $\Pi_1 = \Pi_5$ and 
$\Pi_4 = t_0(R_1{C_0})^{-1}$. From Eqs. (9) - (14) 
one can see that the circuit with a DC voltage corresponds 
to $\Pi_4 = 0$. The initial condition for Eq.(8) is ${I_2}(0) = 1$ 
and  for  Eq. (9) - $U(0) = 1$. Correspondingly, for 
Eqs.(12) - (14) the initial conditions  are  of 
the form: ${I_1}(0) = {I_2}(0) = 0$, $U(0) = 1$. 
To diminish the arbitrariness  in 
choosing initial conditions for Eqs.(1) - (3), 
we take one of the values  on 
the surface of stable solutions of Eqs.(1) - (3) 
at $I_2 = 1 = const$: $X(0) = Y(0) = - 1.1419$, 
$Z(0)= - 0.40048.$ In computer experiments, we set 
$s = {\nu_m}{\nu}^{-1} = 0.3779\cdot10^{-3}$ practically  
in  all  the  cases  (below,  other 
values of $X(0)$, $Y(0)$, $Z(0)$ and $s$ are marked off). 
\par
It can be shown that Eqs.(1) - (3) and (8), and also 
(1) - (3) and  (9) - (11), (1) - (3) 
and (12) - (14), without load resistor, exhibit an  
asymptotic behavior similar to that of 
(50) from [1]. Actually, for  circuit S1 
and  also  for  a  one-loop 
circuit S3, we obtain (below $t_*$ is  the  time,  
corresponding  to  the singularity): 
\begin{eqnarray}
X{\sim}-1.8315(t_* - t)^{-1}, \\
Y{\sim}Z{\sim}-0.9569i(r_1(as{\Pi_2})^{-1})^{1/2}(t_* - t)^{-1},
\\ 
I_2{\sim}-1.9139i(a{\Pi_2}({r_1}s)^{-1})^{1/2}(t_* - t)^{-1/2}.
\end{eqnarray}
\par
Similar asymptotes for a  one-loop  circuit  have  the  following 
form: 
\begin{eqnarray}
X \sim - 1.221(a{\Pi_2}{G_3}{G_4}^{-1})^{1/2}(t_* - t)^{-1}, \\
Y \sim 1.105i(G_3(as{\Pi_2})^{-1})^{1/2}(t_* - t)^{-1}, \\
Z \sim 1.105i{G_3}{G_4}^{-1/2}(t_* - t)^{-1}, 
\end{eqnarray}
\begin{eqnarray}
I_2 \sim 1.105ia{\Pi_2}{r_1}^{-1}(sG_4)^{-1/2}(t_* - t)^{-1}, \\
U \sim 1.105ia{\Pi_2}{\Pi_4}{r_1}^{-1}(sG_4)^{-1/2}{\ln(t_* - t)}, 
\end{eqnarray}
where $G_3  = 1 + \Pi_2$, $G_4 = G_3(a{\Pi_2})^{-1} - a_1$. 
The imaginary unit $i$ in (15) - (22) points to a fluctuating 
character of transient processes in circuits S1 - S3. 
It is worth while noticing that without NE in  circuits 
S1 and S2, a fluctuating process is impossible. 
\par
We noted above that knowing the $X$, $Y$, $Z$ and $I$ 
amplitudes  we  can 
plot  spatial  field  distributions  of 
hydrodynamical  ($u(r,z)$)  and 
current (${u_T}(r,z)$) velocities at any point of  time. 
Therefore  it  is 
necessary to solve Eqs.(4) - (5) and (6) - (7). 
\par
Equations (1) - (3), (4) - (5), (6) - (7) and (8) - (14) 
are dynamic  systems  which 
are nonautonomous in general. Therefore in their study a 
mathematical 
technique of the theory of dynamic systems [3] - 
[7] can by employed. 
Particularly, one of the methods used below is that of Lyapunov's 
characteristic exponents [8], [9]. Despite the 
fact that  Eqs.(1) - (14)  describe 
physical processes in an open dissipative system, 
which,  as  will  be 
shown below, refers to the class of mixing dynamic 
systems  (K-systems [10], [11]), 
determining relaxation to a thermodynamic  equilibrium [11]. 
If instant values of Lyapunov's exponents are regarded as time 
averaged logarithms of Yakobian's eigenvalue modules of a 
linearized 
dynamic system, then these exponents can be used to analyze a  
transient chaos in the dynamic system and a 
time  behavior  of  the  phase 
space fractal dimensionality and also of the Kolmogorov-Sinai metrical 
entropy, related with the positive Lyapunov's exponents 
$\{{\lambda_i}^{+}\}$ by $S = \sum_i{{\lambda_i}^{+}}$ 
[4]. To calculate Lyapunov's exponents we 
use  the  algorithm proposed in [8], [9] and 
to  calculate  fractal  dimensionality  -  the 
Kaplan-Yorke formula [12]: 
$d_L = j + \sum_{i=1}^j{{\lambda_i}|\lambda_{i+1}|^{-1}}$, 
where  $j$  is evaluated from the conditions 
$\sum_{i=1}^j{\lambda_i} > 0$ and 
$\sum_{i=1}^{j+1}{\lambda_i} < 0$.  Besides 
below, parallel with the Kolmogorov-Sinai entropy S, we use the  value 
$F = \sum_i{\lambda_i}$. Computer experiments show that 
in our  dynamic  system  the 
case of all the Lapunov's exponents being positive is realized. 
Here,  by comparing $S$ and $F$, we can find a characteristic 
time when the behavior 
of the process changes, i.e. the so-called explosive regime 
appears. 
\par
A characteristic property of the  dynamic  systems  described  by 
Eqs. (1) - (3) and (8)-(14) is the existence  of  
a  one-to-one  phase  space 
mapping onto the flat $U - I$, which is sometimes called a 
$UI$-characteristics (VCC; It should be noted that, 
strictly speaking,  the  notion 
of VCC is applicable only for a stationary state of the dynamic 
system). With a uniform step of the amplitude mapping in time 
$(\Delta t)$, 
VCC can be regarded as the Poincare point mapping (accurate within the 
step $\Delta t$).  There is a stationary state at a DC 
voltage in the  circuit 
S2, also an analytical expression for VCC can be found for this 
circuit. Actually, the current in NE is determined  by  the  following 
expression: 
\begin{equation}
I_2 = C_1 + {C_2}{r_1}^{-1}Z 
\end{equation}
($C_1 = U{G_2}^{-1}$, $C_2 =
{G_1}{\Pi_2}{a_3}{G_2}^{-1}$), therefore VCC takes the form 
\begin{equation}
U_2 = {\Pi_2}(1 - {a_3}{C_2}^{-1})I_2 + {\Pi_2}{a_3}{C_1}{C_2}^{-1}. 
\end{equation}
A stationary solution for circuit S2 at $\Pi_4 = 0$ 
has the form 
\begin{eqnarray}
X_{\infty} = \pm(-b{Z_\infty}{I_{2\infty}}{a_1}^{-1})^{1/2}, \\
Y_\infty = \pm(-b{Z_\infty}({a_1}{I_{2\infty}})^{-1})^{1/2}, 
\end{eqnarray}
\begin{eqnarray}
Z_\infty & = & -\frac{r_1{C_1}(1-2{a_1}C_1)}{2{C_2}(1-{a_1}C_2)} 
\biggl\{1 + \nonumber\\
& & +\biggl(1 - \frac{4{C_2}(1-{a_1}^2{C_1}^2r_1)
(1-{a_1}C_2)}{{a_1}{C_1}^2{r_1}(1-2{a_1}C_1)^2}\biggr)^{1/2}\biggr\}, 
\end{eqnarray}
where $I_{2\infty}$ is determined by formula (23). 
It is seen  from  (27),  that  
at  negative  root  values  a  stationary  solution  is   nonexistent. 
Expressions (25)  and  (26)  point to  the  existence  of  two 
stationary solutions in the circuit S2 which differ in
$X_\infty$ and/or  $Y_\infty$  signs. 
\section{Computer experiment. Discussion} 
In our computer experiment we specified the following problems: 
\par
{\bf 1.} To find out how initial data influence the dynamics of 
processes in the dynamic systems under study and 
also the role they play in 
reaching and choosing a stationary state from a set of 
permissible stationary states. 
\par
{\bf 2.} To study the spectral dynamics of Lyapunov's exponents and the 
Kolmogorov-Sinai entropy and also the role of  a  transient  chaos  in 
"forgetting"  the initial condition. 
\par
{\bf 3.} To investigate the dynamics of establishing a  limiting  cycle 
in circuit S3 under a DC voltage and its connection with  the  initial 
data. 
\par
{\bf 4.} To find out the difference between physical processes 
occurring in the dynamic systems with a finite energy content 
(circuits  S1 
and S3) and those in circuits under a DC voltage (circuits S2 
and S3). 
\par
{\bf 5.} To analyze the problem of transformation of the time in the 
model (1) and to study the chaos in a "transformed" dynamic 
system. 
\par
{\bf 6.} To determine and to classify the characteristic 
features of a 
spontaneous breaking symmetry and the transition to turbulence 
resistance in the initially laminar electron fluid. 
\par
{\bf 7.} To study the dynamics of NPT  induced  by  an  external  noise 
(determinate or chaotic) and also the influence of frequency  detuning 
on the transient process in the circuit S3. 
\par
Presented below are principal results of our computer  experiment 
and their brief discussion. 
\par
{\bf 3.1.} The dynamics of transient processes has 
been found to be of a 
threshold character, a transition to a steady-state  regime  occurring 
through the sequence of bifurcations at a sufficient level of 
supercriticality. In this case, the transient 
process in circuits S1 and S2 
is an oscillating one and  a  section  with  a  negative  differential 
resistance appears on the VCC. Fig.2 displays VCC in the  circuit  S1. 
One can see that for a critical regime, the system "wishes" to perform 
a phase transition though due to  the  lack  of  supercriticality  the 
initiated perturbations subside and the system  comes  back  into  the 
steady-state, corresponding to zero current in circuit S1. 
\par
{\bf 3.2.} In the circuit S2 under a DC voltage there is a 
stationary 
state determined by expressions (25) - (27). A transition 
to this state occurs 
through a sequence of bifurcations (curves 1 and 2 in 
Fig.3 show 
respectively the change in the current and  the  voltage  drop  across 
NE). Let us note that without NE in the  circuit  S1  the  fluctuating 
process fails. The fluctuations appear due to the  fact  that  NE  are 
actually a dissipative electrodynamic system where  energy  transforms 
step by step from mechanical degrees of freedom to the electromagnetic 
field and vice versa (see [1]). The  stationary  state  corresponds 
practically to a  zero  current  in  NE  the  amplitude  of  which  is 
sufficient to  sustain  this  regime,  the  voltage  drop  across  the 
conductor being non-zero. The NE resistance in  the  stationary  state 
considerably surpasses its initial value. In the next section we  show 
that it is determined by the formation of  vortex  structures  in  the 
electron fluid (by a spontaneous breaking symmetry), that  is  why  it 
can be referred to as a turbulence resistance. 
\par
As it follows from expressions (25) - (27), the VCC 
corresponds  to  two 
stationary states which differ in the $X$ and (or) $Y$ signs. 
The realization of one of the signs is determined by  
$X(0)$  and  (or)  $Y(0)$.  Our 
simulation showed that the initial state of the system with 
the signature $\{- - -\}$ or $\{- + -\}$, $\{- - 0\}$ or 
$\{- + 0\}$ leads to the choice of the 
stationary state with the signature $\{- - -\}$. The initial 
state with $\{+ + -\}$ or $\{+ - -\}$, $\{+ + 0\}$ or $\{+ - 0\}$  
results  in  the  choice  of  the 
stationary state, respectively, with $\{+ + -\}$ (the  symbol  
$"0"$  points 
that $Z(0) = 0$ and, hence, its sign is neutral). It has been also shown 
that the signature of the stationary state, when $X(0) = Z(0) = 0$, is 
determined by the sign $Y(0)$: the signature $\{- - -\}$ 
corresponds to  the 
signature of the initial state $\{0 - 0\}$, and the signature 
$\{0 + 0\}$ - to 
$\{+ + -\}$. 
\par
{\bf 3.3.} A similar condition takes place in the circuit S3 under a DC 
voltage. The presence of reactance elements in the circuit,  i.e.  the 
inductance $L_0$ and $L_{p0}$, brings about the distortion  
of  the  stationary 
state that existed in the circuit S2.  A limit cycle acts as a 
stationary state in the circuit S3 (see  Fig.4). 
\par
{\bf 3.4.} The stationary solution in the circuit S2 and the limit cycle 
in the circuit S3 at  fixed  circuit  parameters  "forget"  about  the 
starting data due to bifurcations in the dynamic system (the  starting 
data influence the time of transition to the stationary  state  rather 
than the stationary state itself). The  character  of  forgetting  the 
starting data is characterized by the time of uncoupling 
the correlations $\tau_c$, which is connected with the 
Kolmogorov-Sinai entropy by  the 
relation ${\tau_c}\sim S^{-1}$ [10], 
[13]; the entropy in its turn  is  determined  by 
the spectrum of the Lyapunov's exponents. Figure 5 shows the entropy time 
alteration for circuit S2; the star symbol $"*"$ marks 
characteristic moments 
of time corresponding to the cross-sections of the conductor passed by 
the plane $r - z$ in Fig.9. The maximal entropy growth rate and,  hence, 
the maximal rate of the descent of $\tau_c$ falls at 
the region of a transient chaos, which 
corresponds to the process when the  system  passes 
through successive bifurcations. Figure 5 shows  the  initial  section 
$S(t)$ where among the points marked by $"*"$ there are two characteristic 
points: the first one corresponds to the time when  all  the  Lyapunov 
exponents become nonzero (positive) and the derivative 
$S(t)$ is discontinued; the second one corresponds to 
the inflection point of the curve $S(t)$. The break 
point corresponds to the  current  0.9  and  the 
inflection point - to 0.1 - 0.15 (see Fig.5c which shows the curve  of 
current in NE). This allows us, as is shown in the third paper of  the 
present series, to formalize the notion of the "start" and  the  "end" 
of commutation while processing the experimental findings. The maximum 
on the entropy curve (Fig.5a)  corresponds  to  a  maximal  chaos.  On 
reaching this chaos the system "settles down" and $S(t)$ tends 
to a certain limit, which, generally speaking, depends 
on the point of  the phase space where we start from at 
the  initial  moment.  Under  fixed 
circuit parameters, as is shown by computer experiment, the stationary 
state corresponds to an infinite set of starting data  (trajectories). 
This  behavior  is  characteristic  of  dynamic  systems  with  mixing 
(K-systems) [4], [10], [11], [13]. 
\par
{\bf 3.5.} In the indicated set of trajectories there is  one  (in  the 
sense of an optimal control [14]) in 
the $\varepsilon$-vicinity, where the Kolmogorov-Sinai entropy is minimal. 
Therefore it can be  called  geodesic. 
We found it by converting the time in the  initial  equations  of  the 
model (it should be noted that we  were  promoted  to  do  so  by  the 
comments in [13] of that the state  
with  all  the  positive  Lyapunov 
exponents corresponds to a repelling center - a repeller 
[4]; in  this case the state at $t = -\infty$ corresponds to  an  
attractor).  We  expected 
that having passed through the region of a transient chaos, the system 
returns to the state close to the initial one. However this assumption 
proved to be wrong. In the dynamic system obtained by  converting 
the time, when it approaches the onset of the region  of  a  transient 
chaos, there appears chaos in the cross-section of the phase transition 
on the $X - Y$ plane (Fig.6). Since NE are included in the  circuit 
with a DC voltage source, the $X$, $Y$, $Z$ and $I$ 
amplitudes tend to infinity 
during a finite time "covering up" all  the  region  of  the  phase 
transition. Starting from any point on this trajectory we come back to 
the stationary state directly along this line (see Fig.7  which  shows 
the change in $X$ (7a), $Y$ (7c) at the back running time, and 
$X$ (7b), $Y$ (7d) at the 
forward time). Starting from any  other  trajectory  point   except 
geodesic, we will also come  to  the  stationary  state  though  along 
another trajectory, which eventually will tend to a geodesic  one  and 
in this sense a geodesic trajectory is an attractor (see Fig. 8). 
\par
{\bf 3.6.} Polarity reversals of the current occurring in  the  dynamic 
system with constant  local  kinetic  coefficients  during  successive 
bifurcations point to the fact that a  spontaneous  breaking  symmetry 
takes place. Figure 9 shows cross-sections of the conductor by 
the $r-z$ plane at characteristic times marked by $"*$" 
on the curve $S(t)$ (see Fig. 5a). Figure 9a presents the 
initiation of a double stationary point of 
the dynamic system "center-saddle" (3) (the moment of the first vortex 
set initiation) and Figure 9b - the second vortex set initiation. From 
the moment of the first vortex set  initiation,  a  formed  separatrix 
surface divides the region occupied by the  current  into  two  parts, 
with a part of the total current being  partly  "intercepted"  by 
vortices. Sites, marked by "*" in Fig. 9a, correspond to the location of 
a Joule heating  source  due  to  the  disturbance  of  the  conductor 
uniformity. These sites  correspond  to  the  so-called  "hot  points" 
revealed in electrically explosive conductors 
[15] (we  will  discuss 
them in detail in the third paper of the present series). 
During a 
further evolution the region occupied by a laminar current is 
"resqueezed" (Figs 9c and 9d show the field of separatrix at the onset 
and at the end of the commutation respectively). 
\par
Thus, the increase in an effective conductor resistance is a 
consequence of the change in the topology of the conductor 
region 
occupied by the current and not an increase in a local specific 
resistance that is caused by small-scale kinetic fluctuations. 
\par
Figure 9e corresponds to the first connection of current in 
reverse polarity. One can see that it is not the  change  of  particle 
rotation in the vortex which the polarity reversal corresponds to  but 
a reconnection of trajectories belonging  to  the  separatrix  between 
neighboring couples of singular points on the external surface of  the 
conductor. Figure 9f shows the field of separatrices for the state  of 
the dynamic system close to a stationary one. It can by seen  that  in 
the stationary state a low current, sufficient to sustain  the  steady 
state in an open dissipative system, is reached. An effective turbulent 
conductor resistance in the stationary state is by two orders  of 
magnitude greater than the initial one. Figure 10 shows the  field  of 
conduction electron trajectories which corresponds to the  time  close 
to that in Fig. 9f. 
\par
Figure 11 shows the field of hydrodynamic particle  trajectories. 
It is easy to notice that the hot points are squeezed between 
hydrodynamic vortex loops, the particles of which move near the 
axis towards each other. It is known 
[16] that vortex loops can move  along 
the loop axis, the direction of their motion coinciding 
with the 
direction of particles near the axis. Hence, vortex loops,  
the  particles 
of which near the axis move towards each other, are attracted, forming 
pairs similar to Cooper pairs in superconductors 
[17], and loops,  the 
particles of which near the axis  move  in  opposite  directions,  are 
repelled. Due to this fact the conductor starts splitting into  pieces 
which have (see [1]) $k_1 = 0.5k_0$ and 
which are topologically equivalent 
to the sphere without an axis. The fact allows us  to  understand  the 
similarity hypothesized  in [1],  that  
in  splitting  the  following 
hierarchy takes place: $k_2 = 2k_0\rightarrow k_3 =
2k_2\rightarrow k_4 = 2k_3$  etc., proceeding 
on the assumption that particles formed as a result of  the  conductor 
splitting "contain" paired vortex loops. 
\par
{\bf 3.7.} Transient processes in circuits with a finite  energy  store 
differ essentially from those in  circuits  under  a  DC  voltage,  in 
particular, the steady state in the  circuit  S3  corresponds  to  the 
condition $X_\infty = Y_\infty = Z_\infty = 
I_{1\infty} = I_{2\infty} = 0$. Figure 12a shows  VCC  in  the 
circuit S3 the parameters of which are chosen so  that  the  transient 
process in a subcritical regime may be close to the aperiodic 
one: $\Pi_0 = \Pi_1 = \Pi_3 = 1$, $\Pi_2 = 0.1$, 
$\Pi_4 = 0.5$, $\Pi_5 = 0.1$; $X(0) = - 1.1419$, $Y(0) = Z(0) = 
0$, $r_1 = 750$. The onset of  a  curve  portion  with  a  negative 
differential resistance of NE  corresponds  to  a  topological 
reconstruction of singular points  of  
the  dynamic  system  determined  by 
Eqs.(1) (see Figs.12b and 12c), in consequence of which a channel  for 
a laminar flow along the conductor axis is formed. On the initial  and 
decaying portions of VCC where the  circuit  current  is  low,  closed 
toroidal structures may be formed  which  are  ellipsoids  or  spheres 
"dressed" on the axis that is a separatrix (Fig 13). 
\par
It should be noted that in  circuits  with  finite  energy  store 
sources the dynamics of the states of current and  NPT  is  determined 
not only by the control parameter $r_1$  but also by the 
initial amplitude 
of the field of hydrodynamic velocity $X$  which depends on the 
prehistory of the process (in particular, on the process of 
storing mechanical defects during the stage of premelting 
and  melting).  Since  we 
deal with an electrodynamic system, the NFT can be realized  only  due 
to $X$ , i.e. due to energy stored in mechanical degrees of  freedom  in 
the previous stages of the process. This idea is supported by 
experiments where the conductor current was interrupted while 
the conductor 
was still melting. For example, in 
[18] the  conductor  was  shown  to 
break down after about $100 {\mu}s$. 
\par
{\bf 3.8.} For the dynamic system with a back running time, 
separatrices 
split the current into three laminar currents the  boundaries  between 
them being formed by separatrices with vortices,  the  area  of  which 
changes chaotically (in time). Figure 14a shows time variation of  the 
Kolmogorov-Sinai entropy, the point on the curve $S(t)$  corresponds  to 
the onset of the  chaos;  Fig. 14b  shows  separatrices  of  conduction 
electron trajectories at the same time. In this case, the current  and 
voltage drop across the conductor increase and the  impedance  in  its 
absolute value has a considerably less value than  in  the  stationary 
state of the circuit S2. 
\par
{\bf 3.9.} Included as a component of circuits S2 and S3, the source of 
the external noise in the form of $E(t) = E \cos({\omega}t + 
\varphi)$ with a determinant or random (uniformly 
distributed  in  the  space  $[0,2\pi]$)  phase, 
enabled investigation of NPT induced by the  external  noise.  In  the 
circuit S2 under a DC voltage,  time  intervals  between  bifurcations 
were shown to increase by the action of a determinate noise. 
In addition, the transient process in the intervals is  
basically  determined 
by an external harmonic e.m.f. (see Fig. 15 exhibiting the  change  in 
current and voltage in NE). The harmonic perturbation can be  seen  to 
destroy the stationary state in the circuit S2 under a DC  voltage  by 
hindering a non-equilibrium phase transition (decreasing the degree of 
supercriticality) and by changing the form of  the  resulting  current 
and voltage pulses. In the intervals between bifurcations the 
existence of topological structures characteristic of sources 
with a  finite 
store of energy is possible and an effective resistance of NE is close 
to the initial one. 
\par
Exposed to a harmonic e.m.f.  with  a  random  phase,  stochastic 
oscillations are brought about in the circuit S2 and their behavior is 
determined  not  only  by  a  random  phase  transition  but  also  by 
bifurcations. In addition, in the intervals  between  the  latter,  an 
effective resistance of NE does not reach its initial value 
(Fig.16). 
\par
{\bf 3.10.} In the circuit S3 under a DC voltage, the stationary  state 
represents a limiting cycle with a known frequency (in our case 
$\omega = 0.4\pi$). It is interesting to study the influence 
of e.m.f.  having  the 
same frequency and the detunigs  of  frequency  on  processes  in  the 
dynamic system. Figures 17, 18  and  19  show  point  mapping  of  the 
dynamic system in the $I_1 - I_2$  plane for different 
time  intervals.  It 
can be seen that when affected by e.m.f with  a  resonance  frequency, 
the transient chaos takes a finite  time  at  the  end  of  which  the 
limiting cycle is formed representing a finite digital set of  points. 
Hence, the fractal dimension is smaller than the dimension of a  phase 
space. The detuning of frequency results in the appearance of  chaotic 
fluctuations which are qualitatively similar to those appearing  under 
the influence of a harmonic e.m.f. with a random phase. 
\section{Conclusion}
\par
The principal result of the present paper is a demonstration that 
large-scale vortex perturbations determine the dynamics of  states  of 
current in plasma-like media.  It  is  important  that  the  conductor 
material is supposed to be incompressible and local kinetic 
coefficients are 
constant. Hence, we have a sufficiently simple  analogy  with  
equilibrium phase transitions, where  a  local  potential  of  interaction 
between particles does not practically influence the dynamics  of  the 
phase transition [19]. The results 
obtained and  discussed  above  are 
mostly of a qualitative character, as they are not applicable from the 
beginning of the space scale  splitting.  Nevertheless,  they  can  be 
applied for  a  qualitative  analysis  of  experiments  on  electrical 
explosion of conductors which are discussed in the third paper  of  our 
series. 
\ack
\par 
We are pleased to thank A I Tebaikin and D A Kargapolov  for  the 
assistance during the computer experiment and the opportunity  to  use 
the computer graphics software developed by them.                      
\references

\numrefjl{[1]}{Volkov~N~B~and~Iskoldsky~A~M 1993}
{J. Physics A: Math. and Gen.}{{\rm (In this volume).}}

\numrefjl{[2]}{Volkov~N~B~and~Iskoldsky~A~M 1990}{JETP Lett.}
{51}{634.}

\numrefbk{[3]}{Anosov D V and Arnold V I (eds) 1988}
{Dynamical Systems 1.~Ordinary Differencial Equations and 
Smooth Dynamical Systems {\rm (Springer: New York).}}

\numrefbk{[4]}{Sinai Ya G (ed.) 1989}{Dynamical Systems 
2.~Ergodic Theory with Applications to Dynamical 
Systems and Statistical Mechanics 
{\rm (Springer: New York).}}

\numrefbk{[5]}{Arnold V I, Varchenko A N and Gusein-Zade S M 1982}
{Singularities of Differentiable Mappings 1.~Classification of 
Critical Points, Caustics and Wave Fronts 
{\rm (Nauka: Moscow) (in Russian).}}

\numrefbk{[6]}{Arnold V I, Varchenko A N and Gusein-Zade S N 1984}
{Singularities of Differentiable Mappings 2. Monodromy and 
Integral Asymptotic {\rm (Nauka: Moscow) (in Russian).}}

\numrefbk{[7]}{Bautin~N~N and Leontovich~E~A 1984}
{Methods and Technique 
of a Qualitative Study of the Dynamic Systems on the Flat 
{\rm (Nauka: Moscow) (in Russian).}}

\numrefjl{[8]}{Benettin~G, Galgani~L and Strelcyn~J~M 1976}
{Phys. Rev. A}{14}{2338.}

\numrefjl{[9]}{Benettin~G, Froeschle~C and Scheidecker~J~P 1979}
{Phys. Rev. A}{19}{2454.}

\numrefbk{[10]}{Zaslavsky~G~M 1984}{Stochasticity on 
the Dynamical Systems}{(Nauka: Moskow) (in Russian).}

\numrefbk{[11]}{Krylov~N~S 1979}{Works on the Foundation 
of Statistical Physics}{(University Press: Princeton).}

\numrefjl{[12]}{Kaplan~J~L and Yorke~J~A 1979}
{Lect. Notes in Math.}{730}{204.}

\numrefbk{[13]}{Zaslavsky~G~M and Sagdeev~R~Z 1988}
{Introduction to the 
Nonlinear Physics}{(Nauka: Moscow) (in Russian).}

\numrefbk{[14]}{Young~L~C 1969}{Lectures on the 
Calculus of Variations and Optimal
Control Theory}{(W B Sounders Company: Philadelphia).}

\numrefjl{[15]}{Baksht~R~B, Datzko~I~M and 
Korostelev~A~Ph 1985}{JTP}{55}{1540 (in Russian).}

\numrefbk{[16]}{Lamb~H 1932}{Hydrodynamics 
{\rm (University Press: Cambridge).}}

\numrefjl{[17]}{Cooper~L~N 1956}{Phys. Rev.}{104}{1189.}

\numrefjl{[18]}{Abramova~K~B, Zlatin~N~A and 
Peregood~B~P 1975}{JETP}{69}{2007 (in Russian).}

\numrefbk{[19]}{Landau~L~D and Lifshitz~E~M 1969}
{Statistical Physics}{(Addison-Wesley: Massachusets).}
\newpage
\figures
\figcaption{Electric circuits: a - circuit S1, b - S2, c - S3. 
$L_0$ - inductance; $C_0$  - capacitance; $R_1$ - 
ballast  resistance; 
$R_2$ - the resistance of a non-linear element, $L_{p0}$ - its 
external inductance; $R_3$  - load impedance. 
Circuit S1 can be easily transformed to the 
DC current source, if 
we formally let $L_0\rightarrow\infty$. 
Similarly, circuits S2 and S3 can be 
easely transformed to 
the DC voltage sources, if we formally 
let $C_0\rightarrow\infty$.}
\figcaption{$UI$-characteristics of a non-linear element in the circuit 
S1: 1~-~subcritical conditions ($r_1 = 5$); 
2~-~critical conditions ($r_1 = 6.34$); 
3~-~supercriticality ($r_1 = 6.5$).}
\figcaption{Characteristics of the transient process in the circuit S2 
with a DC voltage source: 
1~-~current in the non-linear cell; 
2~-~voltage in the non-linear cell. 
In  the  circuit  S2  at  a  constant $R_2$ 
oscillations are impossible. The 
oscillations  shown  in  the 
figure  are  the result of their own internal 
motion in the non-linear element.}
\figcaption{Characteristics of the transition process 
in  the  circuit  S3 with a DC voltage source: 
1~-~the total current; 
2~-~current in the non-linear cell; 
3~-~a load current. 
Adding the $L_0$ and $L_{p0}$ inductances to the 
circuit S2  results in the 
destruction of the stationary state obtained in the 
circuit S2. 
Instead of it, the limit cycle acts in the circuit S3.}
\figcaption{Time variation of the Kolmogorov-Sinai entropy 
$S(t)$ and the current in the non-linear cell for the 
circuit S2  with a  DC voltage source: 
a - $S(t)$; b - the starting section of the curve 
$S(t)$; c - the starting 
section of the current  curve;  the  sign  "*"  marks 
characteristic points for which sections of the conductor  are 
drawn by the plane 
$r - z$ (Fig. 9): the second point corresponds 
to the break of the curve $S(t)$ when all the Lyapunov 
exponents are non-negative ($I_2 = 0.9$: the beginning of 
switching); the third one - to the inflection point $S(t)$ 
($I_2 \cong 0.1 - 0.15$: the end  of  switching). 
The maximum $S(t)$ corresponds to the maximum transient  chaos 
(to  the  minimum time for tripping the correlation 
${\tau_c} \sim S^{-1}$ [13]).}
\figcaption{Mapping of the phase trajectory on the amplitude $X(t)$ 
and $Y(t)$ plane obtained in the circuit S2 
with  a  DC  voltage source 
and the non-linear element in the model of which the time 
sign is replaced by the opposite one (a non-linear element 
with  the back running time).}
\figcaption{Time variation of the $X$ and $Y$ amplitudes in 
the circuit S2 
with a DC voltage source at a back running and a forward time: 
a - X(t) in the circuit with  the  non-linear 
element  at  the back running time; 
b - $X(t)$ in the circuit  with  the  non-linear 
element  at  the forward time; 
c - $Y(t)$ in the circuit with the non-linear element at the 
back running time; 
d - $Y(t)$ in the circuit with the non-linear element at the 
forward time.}
\figcaption{Time variation in the  length  of  the  
radius-vector  of  the amplitudes $W = (\sum{Y_i})^{1/2}$, 
$\{Y_i\} = \{X, Y, Z, I\}$: 
1 - the marked trajectory (see Fig. 7); 
2 - the trajectory from its $\varepsilon$ - vicinity.}
\figcaption{Cross-section of the separatrix  surface  by  the  
$r-z$  plane, which separates finite trajectories  of 
conduction  electrons from infinite ones in the circuit S2 with a DC 
voltage source at times marked by "*" on the curve S(t) (Fig. 5a). 
Sites, marked by "*" in Fig. 9a, correspond to the location of 
a Joule heating source; arrows in Fig. 9a direct the vector field 
velocity of plasma-like medium particles. The state 
corresponding to Fig. 9f is similar to a stationary one.}
\figcaption{Vector field of the current density in the non-linear 
element corresponding to Fig.  9d: 
$t = 26.59$, $I = -0.906\cdot10^{-2}$, $X = -5.1398$, $Y = -2201.38$, 
$Z = - 1797.53$. The solid line  shows  the  cross-section  
of  the  separatrix surface by the $r-z$ plane.}
\figcaption{Vector field of the hydrodynamic velocity.}
\figcaption{$UI$-characteristics  (a)  and  the  cross-section of the 
separatrix surface at the moment of the first vortex 
system initiation (b) and at the beginning of the $UI$-
characteristics with a negative differential resistance (c) 
by "*" in Fig. 12a in the circuit S3 with a limited energy content.}
\figcaption{Vector field of the current density and the cross-
section of the separatrix surface on the decay section of 
the $UI$-characteristics in the circuit S3 with a limited 
energy content.}
\figcaption{Time variation of the Kolmogorov-Sinai entropy in the 
circuit S2 with a DC voltage source and the non-linear element 
at the back running time (a) and the cross-section of the 
separatrix  surface  by the $r-z$ plane at the time 
marked by "*" on the curve $S(t)$ (b).}
\figcaption{Time variation in  the  current  of  the  
non-linear  element 
(a, b) and the voltage across this  element (c,  d)  in  the 
circuit S2 a DC voltage source and with the 
additional source  of an external harmonic noise.}
\figcaption{Time variation in the voltage across the non-linear 
element in the circuit S2 with a DC voltage 
source and with an 
external harmonic voltage source having a random phase 
distributed in the interval $[0,2\pi]$ with the probability 
$w = (2\pi)^{-1}$: a, b, c and d correspond to various time intervals.}
\figcaption{Point mapping of the phase trajectory on the plane 
$I_2 -I_1$ in the circuit S3 with a DC voltage 
source and an external 
harmonic noise voltage source with the frequency $\omega = 0.2\pi$: 
a - $t\in[0, 50]$; b - $t\in[50, 100]$; c - $t\in[100, 150]$; 
d - $t\in[150, 200]$; e - $t\in[200, 250]$; f - $t\in[250,300]$.}
\figcaption{Point mapping of the phase trajectory on the plane 
$I_2 -I_1$ in the circuit S3 with a DC voltage 
source and an external harmonic noise voltage source with the 
frequency equal to the natural frequency ($\omega = 0.4\pi)$: 
a - $t\in[0, 100]$; b - $t\in[100, 200]$; c - $t\in[200, 300]$.}
\figcaption{Point mapping of the phase trajectory on the plane 
$I_2 -I_1$ in the circuit S3 with a DC voltage 
source and an external harmonic noise voltage source with the frequency 
$\omega = 0.6\pi$: a - $t\in[0, 50]$; b - $t\in[50, 100]$; 
c - $t\in[100, 150]$; d - $t\in[150, 200]$; e - $t\in[200, 250]$; 
f - $t\in[250,300]$.}

\end{document}